\documentclass[useAMS,usenatbib,usegraphicx]{mn2e}
\usepackage{times}

\newif\ifAMStwofonts
\AMStwofontstrue

\newcommand{\simlt}{\lower.5ex\hbox{$\; \buildrel < \over \sim \;$}}
\newcommand{\simgt}{\lower.5ex\hbox{$\; \buildrel > \over \sim \;$}}
\newcommand{\be}{\begin{equation}}
\newcommand{\ba}{\begin{eqnarray}}
\newcommand{\ee}{\end{equation}}
\newcommand{\ea}{\end{eqnarray}}
\newcommand{\etal}{{et al.~}}

\title[Close Pairs of Early-type Galaxies]
{Star formation and nuclear
 activity in close pairs of early-type galaxies}
\author[B. Rogers et al.]
{Ben Rogers$^1$, Ignacio Ferreras$^2$\thanks{E-mail: ferreras@star.ucl.ac.uk},
Sugata Kaviraj$^{2,3}$, Anna Pasquali$^4$ and Marc Sarzi$^5$\\
$^1$ Department of Physics, King's College London, Strand, London WC2R 6LS\\
$^2$ Mullard Space Science Laboratory, University College London, 
Holmbury St Mary, Dorking, Surrey RH5 6NT\\
$^3$ Department of Physics, Denys Wilkinson Building, Oxford, OX1 3RH\\
$^4$ Max-Planck-Institut f\"ur Astronomie, K\"onigstuhl 17, D-69117 Heidelberg, Germany\\
$^5$ Centre for Astrophysics Research, University of Hertfordshire, 
College Lane, Hatfield, Herts AL10 9AB\\
}

\begin{document}
\date{MNRAS,{\it Accepted 2009 July 19. Received 2009 July 03; in original form 2009 May 11}}
\pagerange{\pageref{firstpage}--\pageref{lastpage}} \pubyear{2009}
\maketitle
\label{firstpage}

\begin{abstract}
We extract from the Sloan Digital Sky Survey a sample of $347$ systems
involving early type galaxies separated by less than 30~kpc, in
projection, and 500~km/s in radial velocity. These close pairs are
likely progenitors of dry mergers. The (optical) spectra is used to
determine how the interaction affects the star formation history and
nuclear activity of the galaxies. The emission lines (or lack thereof)
are used to classify the sample into AGN, star forming or
quiescent. Increased AGN activity and reduced star formation in
early-type pairs that already appear to be interacting indicate that
the merging process changes the nature of nebular activity, a finding
that is also supported by an increase in AGN luminosity with
decreasing pair separation. Recent star formation is studied on the
absorption line spectra, both through principal component analysis as
well as via a comparison of the spectra with composite stellar
population models. We find that the level of recent star formation in
close pairs is raised relative to a control sample of early-type
galaxies. This excess of residual star formation is found throughout
the sample of close pairs and does not correlate with pair separation
or with visual signs of interaction.  Our findings are consistent with
a scenario whereby the first stage of the encounter (involving the
outer parts of the halos) trigger residual star formation, followed by
a more efficient inflow towards the centre -- switching to an AGN
phase -- after which the systems are quiescent.
\end{abstract}


\section{Introduction}

It is well known that early type galaxies are dominated by old stellar
populations \cite[see e.g.][]{ka97,sed98}. It has also been shown
through studies of NUV photometry from GALEX \citep{yi05,kav07} as
well as statistical disections of the optical spectra
\citep{pcai,pca,nol07} that a large fraction of early types have
undergone small amounts of recent star formation.  However there is
less certainty as to the cause of this recent star formation. One
possible scenario involves minor mergers from small blobs of gas
surrounding the galaxy \citep{kav08,kav09}. In that case, the young stellar
mass content will be roughly independent of galaxy mass, which implies
recent star formation will be more readily detectable in lower mass
early-types, as observed. This scenario will also result in enhanced
recent star formation within close pairs, where these small pockets of
gas could be disrupted during the encounter. In this paper we study
differences in the spectroscopic data of close pairs involving only
early-type galaxies with the aim of understanding the connection
between galaxy interactions and star formation or AGN
activity. Restricting the selection to pairs involving only
early-types (i.e. precursors of dry mergers) results in a cleaner
sample, minimising the contamination from gas in the interaction process.

Within the standard framework of a $\Lambda$CDM cosmology, elliptical
galaxies are formed through the merger of many smaller systems
\citep[see e.g.][]{dell06} and although the build up of the red
population occurs mainly at high redshift \citep[e.g.][]{bun05,if09},
signs of previous merging events are found in early-type systems, such
as kinematically decoupled cores \citep{dav01,mcd06}, distorted
morphologies, shells and other fine structure \citep{dok05}.

\begin{figure*}
\begin{center}
\includegraphics[width=5in]{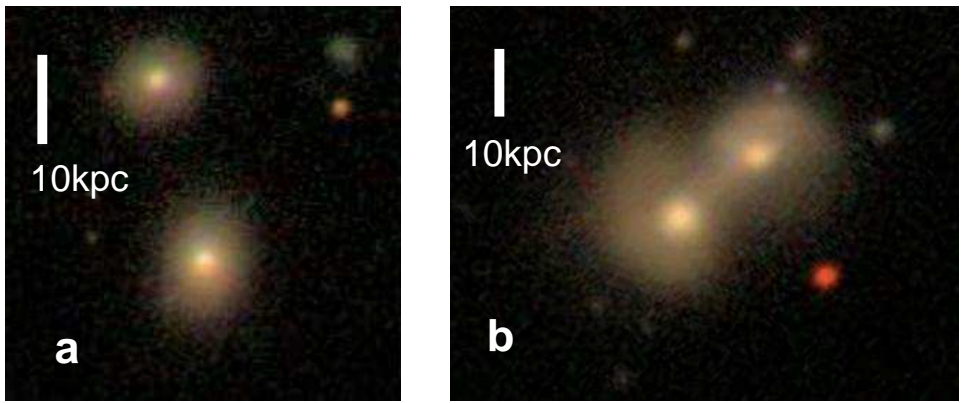}
\end{center}
\caption{An example of close pairs of SDSS early-type galaxies
usewd in this paper:
Non-interacting (a; SDSS J101434.27-005013.1; z=0.045)
vs. interacting (b; SDSS J083645.89+472210.1; z=0.053). 
The vertical bar represents a projected distance of 10~kpc.}
\label{fig:images}
\end{figure*}

The overall old stellar populations found in elliptical galaxies can
be used to constrain the time since the last major episode of star
formation. If the last merger involved late type galaxies, it must
have taken place at early times, otherwise the resulting young
populations would be seen in a photometric/spectroscopic analysis.
However, if the last merger involves gas poor galaxies, one could
still accommodate mergers at late times. It has been shown that models
describing the formation of massive elliptical galaxies require a
non\-dissipative merger (i.e. limited amounts of cold gas) to recreate
the dynamics of the largest ellipticals, such as boxy isophotes and
minimal rotation \citep{naab}, \citep[although see ][]{kang}.
Therefore it is reasonable to assume that the last merger for a
significant fraction of ellipticals involved early-type galaxies
as progenitors \citep{khoc03}.

Given the large fraction of elliptical galaxies likely to contain
small but significant amounts of cold gas \citep{knap89,Yng05},
disruption of this gas via gravitational interactions/harassment or
through 'dry' mergers may well result in small amounts of star
formation. It is also worth considering that previous work
\citep{pcai,pca} has indicated that recent star formation may be more
significant in medium density environments, implying that galaxy
harassment may be efficient at stimulating small episodes of star
formation.

In this paper we investigate these possibilities by looking at {\sl
elliptical only} close pairs.  We analyse the emission spectra to
identify active galaxies (AGN or star forming) and explore whether the
impending merger affects the nature of the activity. We also perform a
comprehensive analysis of the stellar populations, using two
independent methods: principal component analysis, and a maximum
likelihood analysis involving a grid of models. The latter uses a
number of age and metallicity sensitive spectral features measured via
a new estimator of equivalent width that minimises the contamination
from neighbouring lines \citep{bmc}.

\begin{figure}
\begin{center}
\includegraphics[width=3.5in]{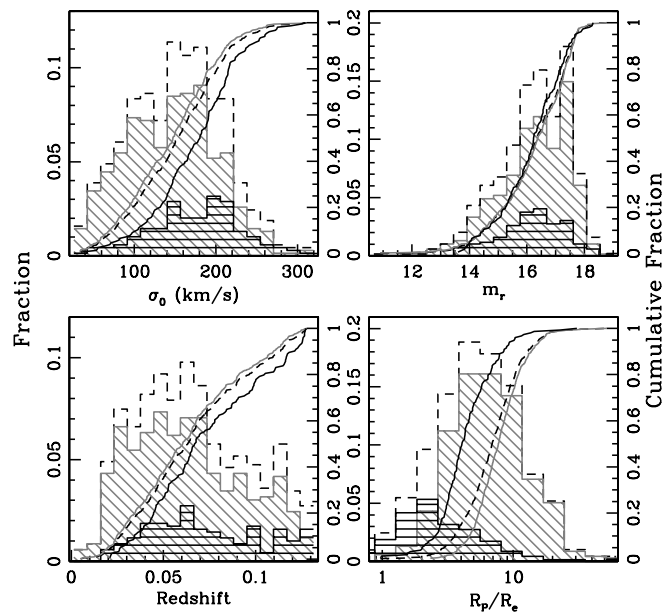}
\end{center}
\caption{The distribution of the sample is shown in terms of redshift
({\sl bottom left}), central velocity dispersion, $\sigma_0$ ({\sl top left}),
apparent r-band magnitude ({\sl top right}) and the ratio between the
projected separation of the pair and the half-light radius,
R$_{\rm P}$/R$_e$ ({\sl bottom right}). The histograms show the distributions of
the pairs depending on whether the system is visually classified as 
interacting (black solid) or not (grey). The full sample corresponds to the black
dashed lines.  Overplotted are the cumulative distributions (with the
same colour coding) which highlight differences between the two types
of pairs.}
\label{fig:sample}
\end{figure}

\section{The Close Pairs Sample}

Our classification starts with an initial sample of $\sim~$3000
galaxies in close pair systems from the Sloan Digital Sky Survey
(SDSS) DR6 database \citep{sdss}. A close pair is defined as a system of
two or more galaxies within a projected physical distance of 30~kpc
and with relative velocity below 500~km/s (i.e. $\Delta$z = 0.0017).
The choice is motivated by previous work on the presence of
interaction signs in close pairs of SDSS galaxies \citep[see e.g. ][]{patt00}.

We make a visual classification of this sample using r-band FITS
stamps retrieved from the SDSS pipeline. The morphology of each system
of galaxies was estimated by eye by B.~R., I.~F., S.~K. and A.~P. In
order to simplify the classification method, we bin the systems into
three types: non-early type, lenticular or elliptical. We emphasize
that our sample targets systems where {\sl both} galaxies are
early-types.  Non-early type systems are those which show one or more
galaxies with structure characteristic of a late-type: a dominant disk
component or spiral arms. Lenticulars and ellipticals are defined as
bulge-dominated galaxies with or without a visible disk component,
respectively.  This classification resulted in a sample of 346
early-type only pairs and one triplet system, comprising 695 galaxies
in total.

These pairs were further divided into categories, based on the
estimated level of interaction occuring within the system. This scheme
is similar to previous classifications \citep[see][]{alo07}, although
we only consider two cases. Our systems are either ``non interacting'' or
``interacting'' depending on whether the close pair shows tidal tails
or other distortions, indicating a significant level of
interaction. Figure~\ref{fig:images} shows an example of each type.
This criterion is also determined by a majority selection
from the four classifiers. Systems for which there was an equal
split between classifiers was considered as ``non interacting''. 

\begin{figure}
\begin{center}
\includegraphics[width=3.5in]{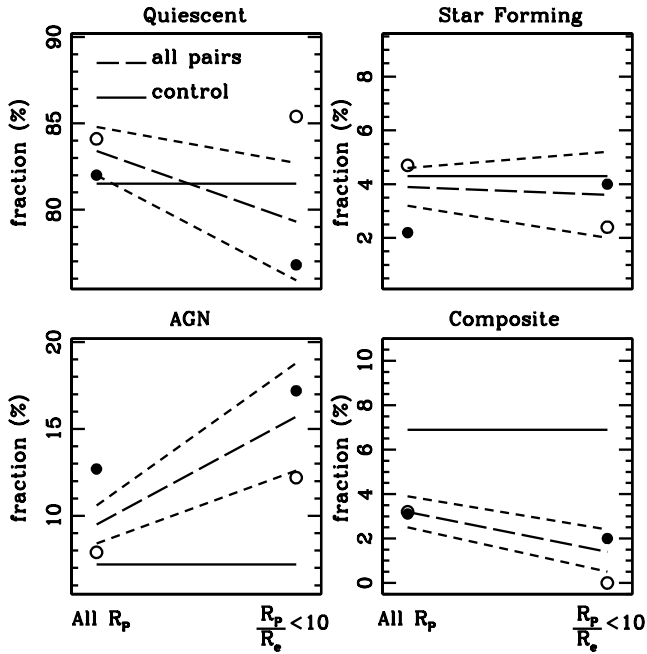}
\end{center}
\caption{The variation of emission line activity with respect to pair
separation is shown for our sample of close pairs (dashed lines,
including Poissonian error bars) and for a control sample of SDSS
early-type galaxies from \citet[][solid lines]{schw07}. For each
panel, the left (right) side of the box corresponds to all galaxies
(galaxies separated from the other member by less than 10 times the
half-light radius). The filled (open) circles correspond to galaxies
visually classified as interacting (non-interacting). This figure
illustrates the fractions shown in table~\ref{tab:sample}.  }
\label{fig:figtab}
\end{figure}

Optical spectra for the entire sample were retrieved from the SDSS DR6
and processed using the PPXF and GANDALF codes \citep{capem,srz},
adapted for fitting SDSS data. These codes provide stellar kinematics
and emission line luminosities by simultaneously fitting both stellar
population models \citep{trem04} and gaussian emission line templates.
The latter -- when present -- allow us to determine the nature of the
ionising source.  A comparison of the Balmer decrement
(H$\alpha$/H$\beta$) to standard case B recombination \citep{osto} is
used to correct for the effect of internal dust by using the
extinction law of \citet{cal}.  The emission spectra is removed from
the observed data to determine the original SED of the underlying
stellar populations. This is an important point when comparing the
equivalent widths of absorption features with synthetic models in
order to constrain the star formation history (SFH). 

Figure~\ref{fig:sample} shows the general properties of our early-type
close pairs sample. The histograms are coded with respect to the
interaction type: non-interacting systems are shown in grey, and
interacting pairs are shown in black. The full sample is shown as a
dashed histogram. We include in the figure the cumulative
distributions to help assess the difference between samples. We draw
attention to the central velocity dispersion ($\sigma_0$; top-left),
whose distribution is slightly biased towards higher values for
interacting galaxies, as expected \citep{park09}. We note that a
similar trend was found by \citet{alo07}, who showed that the
estimated black hole mass, derived from the $\sigma_0$-BH mass
relation, was systematically higher for interacting pairs. We find no
difference with respect to interaction in the distribution of absolute
magnitude ({\sl top-right}) or redshift ({\sl bottom-left}). While one
could expect a bias in the classification, such that galaxies with
brighter apparent magnitudes would have more visible debris, therefore
increasing the fraction of ``interacting'' types, the distribution of
apparent r-band magnitude shows no such trend ({\sl top-right}). The
distribution of interacting galaxies appears clearly biased towards
smaller separations (defined as the ratio between pair separation and
the half-light radius of the galaxy: R$_{\rm P}$/R$_e$;
bottom-right).

\section{Comparing the Emission Line Spectra}

Mergers and interactions disturb the gas in galaxies, and may lead to
both AGN activity as the gas is driven towards the central
super-massive black hole, or star formation if gas clouds can cool and
collapse. Thus, identifying those galaxies which have undergone such
activity in our sample will allow us to gauge the efficiency of this
process in the specific case of gas-poor encounters between early-type
galaxies.  The identification of the ionising source in a galaxy is
possible using emission line diagnostic diagrams \citep[or BPT
diagrams, see ][]{bpt}.  In this paper we use the standard BPT diagram
juxtaposing the $[$NII$]/$H$\alpha$ and $[$OIII$]/$H$\beta$ line
ratios \citep{vo87}. The classification scheme of galaxies according
to this diagram is fully explained in \citet{kew06} and is summarised
below.

Star-forming galaxies are found below the
\citet{kauf03} line: 

\be \log([{\rm OIII}]/{\rm H}\beta) =
\frac{0.61}{\log([{\rm NII}]/{\rm H}\alpha)-0.05} + 1.3.  \ee

Galaxies above this line in the BPT diagram are likely to contain
emission lines from an AGN. These galaxies are subdivided into two
further sets: 'Composite' galaxies, possibly containing both
an AGN and star forming regions, and galaxies whose ionisation spectra
is dominated by AGN activity.  Composite objects are those galaxies
which lie above the \citet{kauf03} star formation line but below the
theoretical upper limit of emission stimulated by star formation,
namely \citep{kew01}: 

\be \log([{\rm OIII}]/{\rm
H}\beta)=\frac{0.61}{\log([{\rm NII}]/{\rm H}\alpha)-0.47}+1.19.  \ee

Galaxies above this upper limit are said to have their emission
dominated by an AGN. While it is possible to continue splitting the
AGN population into Seyferts and LINERs, this adds little to our
analysis, so it is omitted.  Note that for a reliable classification,
the four emission lines involved are required to have S/N $\geq$ 3. We
identify $\sim$ 20\% of our close pairs sample ($\sim$ 140 galaxies)
as 'active' galaxies. 

\begin{figure}
\begin{center}
\includegraphics[width=3.5in]{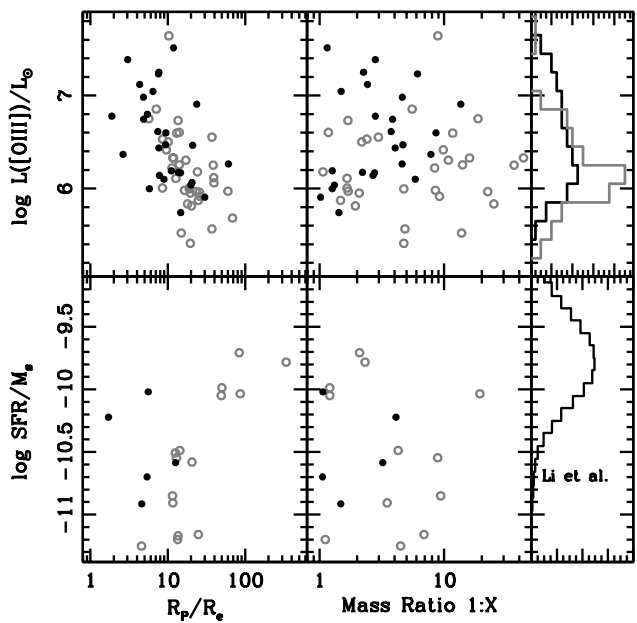}
\end{center}
\caption{{\sl Top:} The luminosity of the [OIII] (5007\AA ) line is shown with respect to
the ratio between projected separation and half-light radius, R$_{\rm P}$/R$_e$
({\sl left}), and stellar mass ratio ({\sl right}) of the Seyfert, LINER and
unclassified AGN galaxies. Galaxies in pairs classified as interacting
(non-interacting) are shown as black solid (open grey) circles, respectively.
The histograms are colour coded in the same way.
{\sl Bottom:} Specific star formation rates with respect to the ratio
between projected separation and half-light radius (R$_{\rm P}$/R$_e$,
{\sl left}), and stellar mass ratio ({\sl right}) for the star forming
sub-sample.  Galaxies in pairs classified as interacting
(non-interacting) are shown as black solid (open grey) circles,
respectively. The histogram is the distribution of specific star 
formation rates from the general sample of SDSS galaxies from
\citet{li08}.}
\label{fig:agnSF}
\end{figure}

\begin{table*}
\caption{Emission line classification for close pairs}
\label{tab:sample}
\begin{tabular}{lccccccc}
      Sample & \multicolumn{2}{c}{Close Pairs} & \multicolumn{2}{c}{Interacting} & 
      \multicolumn{2}{c}{Non interacting} & Comparison$^1$ \\
         & ALL  & R$_{\rm P}/R_e<10$ &
      ALL & R$_{\rm P}/R_e<10$ &
      ALL & R$_{\rm P}/R_e<10$\\
      \hline
      Number       & 695    & 140    & 467    &  41    & 228 & 99 &  15729   \\
      Quiescent    & 83.4\% & 79.3\% & 82.0\% & 76.8\% & 84.1\% & 85.4\% & 81.5\%\\
      Star Forming &  3.9\% &  3.6\% &  2.2\% &  4.0\% &  4.7\% &  2.4\% &  4.3\%\\
      Composite    &  3.2\% &  1.4\% &  3.1\% &  2.0\% &  3.2\% & ---    &  6.9\%\\
      AGN          &  9.5\% & 15.7\% & 12.7\% & 17.2\% &  7.9\% & 12.2\% &  7.2\%\\
      \hline
    \end{tabular}

{$^1$ The comparison sample is taken from \citet{schw07}.\hfill}
\end{table*}

\subsection{Frequency of Active Galaxies}

Shown in table ~\ref{tab:sample} are the fractions of the sample with
signs of AGN activity as a function of the pair interaction type. For
a control sample of early-type galaxies we cannot use our original set
of early-type galaxies \citep{pca}.  That sample was extracted from
\citet{bern06}, which incorporates a colour cut in the selection,
resulting in a bias on the number of active galaxies. Therefore we
make use of a more general data set comprising morphologically
classified early-type galaxies from \citet{schw07}, taken from SDSS
DR4 and spaninng a similar range in redshift. Figure~\ref{fig:figtab}
shows the dependence of emission line activity with respect to pair
separation (R$_{\rm P}$).  The dashed lines illustrate the difference
between the full sample (left side of each panel) and a subsample of
close pairs, defined as those galaxies for which R$_{\rm P}<10$R$_e$
(right side of each panel).  Short dashed lines show the Poissonian
error bars. The control sample of \citet{schw07} is given by a
horizontal solid line in each panel.  We also illustrate the
dependence on signs of interaction: filled (open) circles correspond
to galaxies classified as interacting (non-interacting).  One can see
that the differences between early-type galaxies in close pairs and in
the control sample is quite small, but it appears to be more
significant in systems with visual signs of interaction.

Previous studies have found an increased level of both star formation
\citep{lam03,alo04,woo06,knap09} and AGN activity \citep{alo07,keel} within
close pair samples not segregated with respect to morphology. However,
table~\ref{tab:sample} shows that restricting the analysis to
early-type galaxies results in little or no significant difference --
within Poissonian fluctuations -- in activity between close pairs and
the general population. This effect may be due to the lack of a
substantial amount of gas in early-type galaxies to fuel either AGN
activity or star formation. The small but non-negligible fraction of
'active' galaxies found both in a general and a close pairs sample
could suggest that this activity is caused by a mechanism other than
major mergers. A minor merger scenario \citep{kav09} could explain
those fractions and the insensitivity to being in a close pair.

On the other hand, while the emission line properties of the overall
close pair sample is consistent with a general sample, galaxies with
visual signs of interaction show a higher fraction of AGN compared to
both \citet{schw07} and our non-interacting subsample. This result is
significant within Poisson error bars.  We note that the observed
increase is in agreement with both \citet{alo07}, involving close
pairs from SDSS, and \citet{lirgs}, who targeted Luminous InfraRed
Galaxies (LIRGs). The underlying fractions in both samples are clearly
dependent on the type of galaxy targeted, but the increase is
consistent.  The other point to notice is that the close pair
early-type galaxies with visual signs of interaction also feature a
lack of star formation. This is in contrast to previous studies on
close pairs of all morphological types, which find a strong link
between decreasing separation and increased star formation \citep[see
e.g.][]{alo04,woo06,li08}. The interesting aspect is that the decrease
in the fraction of star forming galaxies is coincident with an
increase in the AGN fraction, a topic that will be discussed later
on. Furthermore, the number of galaxies in the transition region
between AGN and star formation ({\sl bottom-right} panel of
figure~\ref{fig:figtab}) decreases at small pair
separation (R$_{\rm P}<10$R$_e$). Although speculative, one could
envision those trends as an increase in AGN activity with decreasing
pair separation.

Alternatively, it is possible that the signature of an increased
black-hole accretion triggered by galactic interactions is damped by
the presence of diffuse LINER-like emission that can masquerade as AGN
activity, as recently suggested by \citet{stat08}. This would be the
case, in particular, if the power of such diffuse emission is limited
by the amount of ionising photons produced by the old stellar
populations that can reach the gas in early-type galaxies, rather than
by the amount of such material.

\begin{figure}
\begin{center}
\includegraphics[width=3.5in]{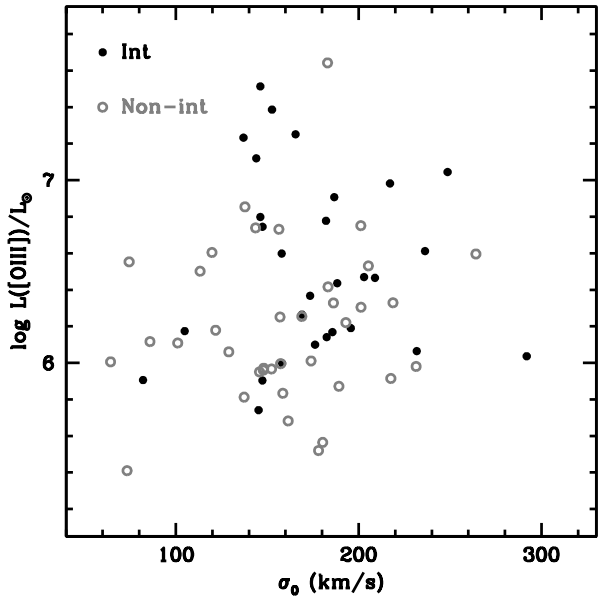}
\end{center}
\caption{The luminosity of the [OIII] line as 
a function of velocity dispersion. The sample is 
segregated into interacting (black) and 
noninteracting (grey). We note that the interacting 
sample has an increased average luminosity but also 
that the correlation with mass is very weak for this 
sample. 
\label{fig:LOIII}}
\end{figure}

\subsection{Star Formation vs. AGN activity}

To determine the possible effects of an increasing inflow of material
into the centres of early-type galaxies in close pairs, we compare the
star formation rate and the AGN activity from the luminosity of
targeted emission lines. One would expect those to depend on the
relative velocity, mass ratio or separation between galaxies. The star
formation rate (SFR) is estimated from the luminosity of the H$\alpha$
emission line, using the standard correlation \citep{ken98}. In order to
factor out the stellar mass, one can also define the specific star
formation rate (i.e. the SFR per unit stellar mass) taking the stellar
masses from the Garching catalogues \citep{gal05}. The AGN activity
can be traced using the luminosity of the [OIII] 5007\AA\ line which
scales with the bolometric luminosity of the AGN and thus the
accretion rate of the the central super-massive black hole \citep{hck04}.

In figure~\ref{fig:agnSF} ({\sl top}) we show the luminosity of the
[OIII] line as a function of the separation ({\sl left}) and the
stellar mass ratio between the members of the pair ({\sl right}). We
only show those galaxies classified as pure AGN, since massive stars
will also affect the luminosity of the [OIII] line, contaminating the
interpretation of L([OIII]). The accretion rate is found to correlate
with separation. The correlation holds even if we consider only the
non-interacting pairs (grey open circles), which are overall found at
separations $\simgt 10$R$_e$. This suggests that even in the absence
of visual signs of interaction one could detect the effects of a close
pair interaction from the activity of the central super-massive black
hole. On the other hand, L[OIII] does not correlate with the stellar
mass ratio ({\sl right}). The vertical histogram ({\sl top-right})
illustrates the trend between visual signs of interaction and the
activity of the central nucleus.

It is important to notice that this correlation does not reflect
any bias regarding stellar mass. Shown in figure~\ref{fig:LOIII} is
the luminosity of the [OIII] line as a function of velocity dispersion
(i.e. a proxy for mass), we see that the increase of L[OIII] due to
the mass of the galaxy is minimal compared to the correlation seen in
figure~\ref{fig:agnSF} with respect to pair separation.

\begin{figure}
\begin{center}
\includegraphics[width=3.5in]{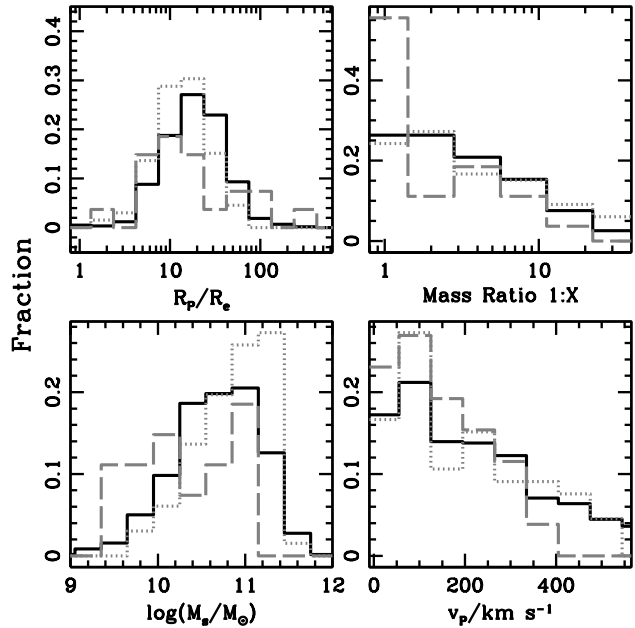}
\end{center}
\caption{Sample properties when segregated with respect
to emission line activity. The solid black, dashed grey and
dotted grey histograms correspond to quiescent, star-forming
and AGN galaxies, respectively.
\label{fig:sample2}}
\end{figure}

As shown in table~\ref{tab:sample}, only 4\% of the total sample have
emission lines consistent with star formation. Those galaxies are
shown in the bottom panels of figure~\ref{fig:agnSF}, where we present
the specific star formation rate with respect to projected separation
({\sl left}) or mass ratio ({\sl right}). On the far right panel, the
distribution of the specific star formation rate for the general
sample of SDSS galaxies from \citet{li08} is given for comparison. As
expected, our early-type close pairs sit at the low end of the
distribution for a general sample. No significant trend is seen,
although one could glimpse a correlation in the sample of
non-interacting pairs (open grey circles) such that the specific star
formation decreases with decreasing separation. Although weak, this
trend fits with the increasing L([OIII]) as the pairs get closer,
suggesting a transition from on-going star formation towards AGN
activity.

Figure~\ref{fig:sample2} shows the distribution of some properties of
the sample when segregating with respect to emission line activity.
The solid black, dashed grey and dotted grey histograms correspond to
quiescent, star forming and AGN galaxies, respectively. Given the
large differences in the number of galaxies belonging to each class,
we show there histograms normalized to the number of members in each
class.  It is worth noting the strong correlation with respect to
galaxy mass ({\sl bottom-left}), with the AGN and star forming galaxies
dominating the top and bottom ends in stellar mass, respectively. The
characteristic mass of a quiescent galaxy sits in between these
two. Regarding mass ratios ({\sl top-right}), we find a significant enhancement
of star formation for more equal (but low-mass) ratios. This enhancement
is also seen for lower relative velocities (defined as the
difference between radial velocities; {\sl bottom-right}).

\begin{figure}
\begin{center}
\includegraphics[width=3.5in]{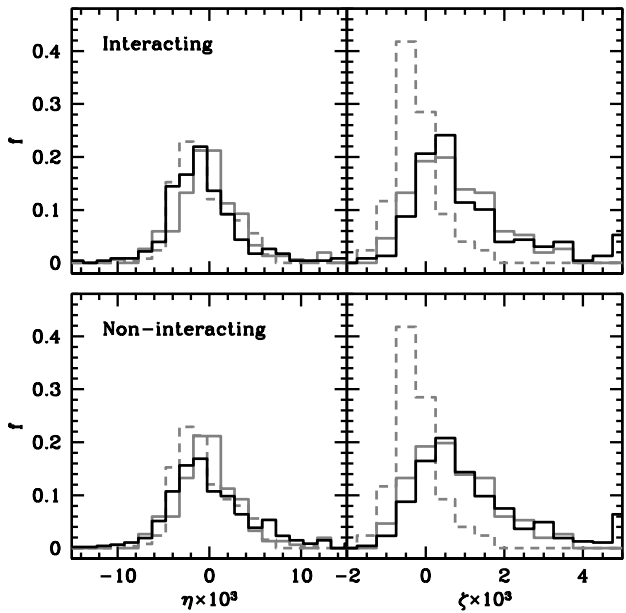}
\end{center}
\caption{The distribution of the sample within $\eta$ and $\zeta$ PCA
space. The close pair sample (black histograms) is separated with
respect to the visual classification: non-interacting ({\sl bottom}) and
interacting ({\sl top}). For comparison, SDSS early-type galaxies with a GALEX
detection are shown as grey histograms, with the dashed (solid) lines
corresponding to red (blue) galaxies. The colour separation is based
on the criterion for the presence of recent star formation,
namely NUV$-$r$<$4.9 for blue galaxies and NUV$-$r$>$5.9 for red galaxies
\citep[see e.g.][]{kav07,pca}.}
\label{fig:pca1}
\end{figure}

\section{Comparing Absorption Line Spectra}

In addition to AGN activity and ongoing star formation -- which are
processes triggered on shorter timescales -- we also explore the
effect of the interaction on the stellar populations as seen from the
continuum and the absorption line spectra. In principle, one may not
expect any significant changes in the properties of the bulk of the
stellar populations during the first stages of an interaction
(i.e. the phase we are only sensitive to in this sample). However,
small and recent episodes of star formation can be detected in the
optical spectra as shown in otherwise red and dead early-type galaxies
\citep{pca}. Galaxies with recent but no on-going star formation will
be excluded from the analysis in the previous section, based on the
emission line spectra. Hence, by extending our analysis to the
continuum and the absorption lines, we increase our sensitivity to
detecting the effect of a close encounter between early-type
galaxies. We consider two independent methods to determine differences
in the stellar populations. The first one involves
Principal Component Analysis (PCA), a model-independent technique
aimed at extracting ``directions'' in a vector space spanned by the
spectral data, along which the variation is maximal. A more
conventional second approach targets age-sensitive spectral features
and compares them with a grid of models combined with population
synthesis spectra to constrain the star formation histories \citep[see
e.g.][]{bmc}.

\subsection{Principal Component Analysis (PCA)}

In order to maximally extract information from spectral data and to
identify the smallest differences between them, we apply the method of
Principal Component Analysis. It has proven to be a useful tool in the
analysis of extremely homogeneous samples, as shown in \citet{pcai} and
\citet{pca}, where it was possible to identify small amounts of recent
star formation at the level of a few percent in mass of $\sim$1~Gyr
old stars from the optical spectra.

\begin{figure}
\begin{center}
\includegraphics[width=3.5in]{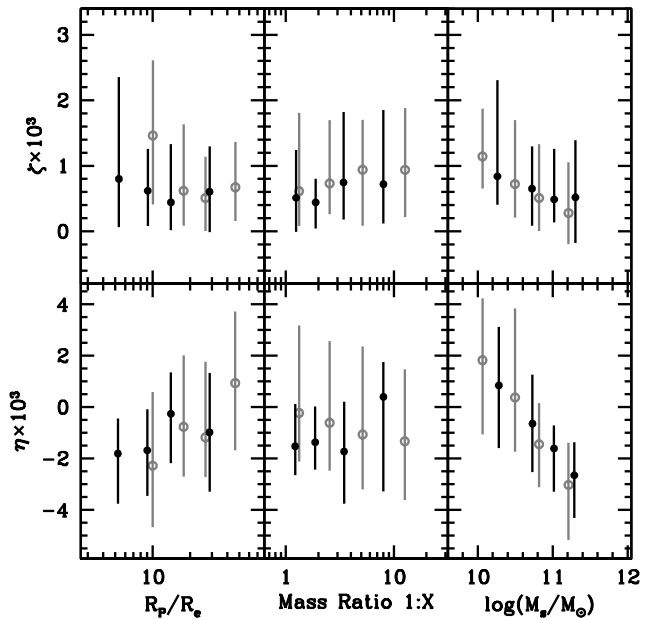}
\end{center}
\caption{The parameters $\eta$ ({\sl bottom}) and $\zeta$ ({\sl top}) derived 
from PCA are plotted as a function of the ratio between
projected separation and half-light radius ({\sl left}); 
mass ratio ({\sl centre}) and stellar mass ({\sl right}). The
black solid (grey open) circles are averages for the
pairs with (without) visual signs of interaction.
The vertical lines span the $25^{\rm th}-75^{\rm th}$ percentiles
in each bin.}
\label{fig:pca2}
\end{figure}

We take the principal components (i.e. the basis ``spectral vectors''
on which the SEDs of these galaxies are projected) from a general
(i.e. non close-pairs) sample of $\sim$7000 early-type galaxies
extracted from SDSS, previously defined and analysed in
\citet{pca}. We refer the interested reader to that paper for details
of the method. This comparison set is a volume-limited sample
extracted from the larger set of SDSS early-type galaxies defined in
\citet{bern06}, with the constraints: M$_r\leq-$21 and z$\leq$0.1,
with a further constraint on the signal-to-noise ratio of the spectra,
S/N$\geq$ 15 per pixel. This control sample provides us with the
pre-processed basis spectra for the analysis (i.e. the principal
components). After de-redshifting and correcting for Galaxy dust
absorption, the spectra from the early-type galaxies in close pairs
are then projected onto these eigenvectors, giving the
projected components (i.e. PC1, PC2) that quantify the relative
weight of each eigenvector in the construction of the spectra:

\begin{figure*}
\begin{center}
\includegraphics[width=5in]{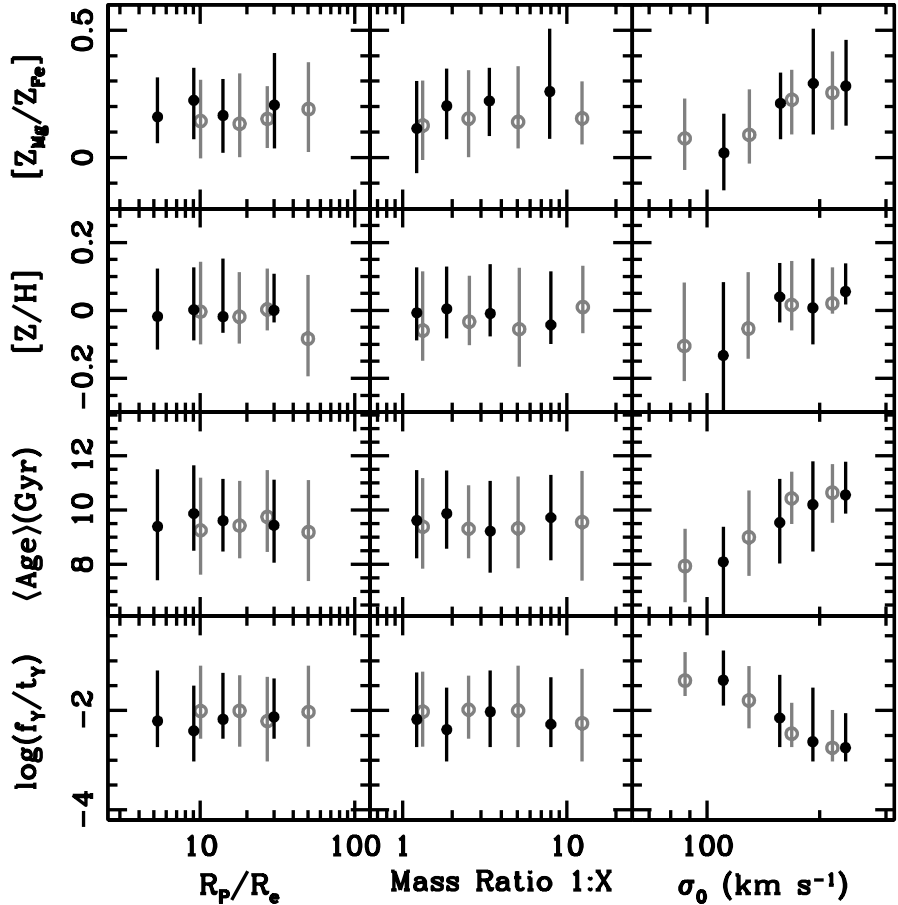}
\end{center}
\caption{Best-fit stellar populations for a 2-burst
model. The analysis involves a number of age- and
metal-sensitive line strengths (see text for details).
From top to bottom, the Mg-to-Fe ratio, metallicity,
average age and the ratio between the young mass fraction and
its age, are shown with respect to projected separation ({\sl left}),
mass ratio ({\sl centre}) and central velocity dispersion ({\sl right}).
Solid black (open grey) circles correspond to galaxies 
with (without) visual signs of interaction. The vertical
lines span the $25^{\rm th}-75^{\rm th}$ percentiles of 
the distribution within each bin.}
\label{fig:2bst}
\end{figure*}

\be 
{\rm PC}1_i =
\vec{\Phi}_i\cdot\vec{e}_1=\sum_{j=1}^N \Phi_i(\lambda_j)e_1(\lambda_j), 
\label{eq:proj}
\ee 

\noindent
and similarly for PC2, PC3, etc... While it is possible to determine
the principal components straight from the close pairs data set, the
smaller sample size means the extracted eigenvectors will be less
robust. PCA relies on the variance of the data set to define an
eigenvector and so large sets are preferred.

\subsubsection{$\eta$ and $\zeta$ components}

The most successful mapping of PCA to extract information from the
underlying stellar populations was found through the first two
principal components. As shown in \citet{pca} these two components, PC1
and PC2, are consistent with an old and young stellar
population, respectively. The eigenvector of PC1 shows a pronounced
4000\AA\ break, significant metal absorption (e.g. H \& K Ca~II lines
or a prominent G Band) and little or no Balmer absorption. In
contrast, the eigenvector of PC2 features a blue continuum with
well-defined Balmer lines.  The positive correlation between PC1 and
PC2 \citep[see figure~3 of ][]{pca}, is likely caused by the relative
values required to reconstruct the spectra (e.g. the shape of the
continuum), which depends on the age and metallicity of the stars. 
Relative to this relationship between PC1 and PC2, a
galaxy with a higher value of PC2 suggests the presence of a young
sub-population. This idea was confirmed through a two-component stellar
population model, in which the excess of the PC1-PC2 relationship was
found to correlate with the mass fraction in young stars.

We also found a consistent correlation between the projections of the
principal components and NUV photometry from GALEX. The NUV spectral
region ($\lambda\sim 2300$\AA\ ) is very sensitive to the presence of
small fractions of young stars. The (NUV$-$r) colour has been shown to
serve as an excellent indicator of recent star formation
\citep{schw07,kav07}. We used GALEX photometry to define two subsets
of galaxies, the first one is NUV bright (NUV$-$r$\leq$4.9) and
represents galaxies that have undergone recent star formation (within
$\sim$1Gyr). The second one is NUV faint (NUV$-$r$\geq$5.9) and
corresponds to an old, quiescent population. \citet{pca} showed that
the projections of the principal components -- {\sl which only use the
optical spectra} -- could be used to discriminate between these two
populations of galaxies.  Thus from the linear fit of the PC1 v. PC2
correlation, we define $\eta$ as the distance along the relationship
(sensitive to average age and metallicity) and $\zeta$ as the residual
from this fit (sensitive to recent star formation).

\subsubsection{Results}

Figure~\ref{fig:pca1} shows the distribution of components $\eta$
({\sl left}) and $\zeta$ ({\sl right}) for the close pairs (solid line
histograms). The top ({\sl bottom}) panels correspond to the interacting
(non-interacting) visual type, respectively.  In order to compare with
a 'control' sample, we also show in each panel the histograms for the
sample of (non-close pair) elliptical galaxies \citep{pca}, segregated
with respect to NUV$-$r colour. NUV bright (faint) galaxies are shown
as grey solid (dashed) lines. Both ``interacting'' and ``non-interacting''
galaxies have the same distribution of the $\eta$ and $\zeta$
components.  Thus, PCA indicates that there is little difference
between the average populations of the two classes, which suggests
that the visual level of disruption does not dictate the amount of
recent star formation. Given these results it is not obvious that the
interaction plays a significant role in shaping the stellar
populations.  This might be expected since our systems are the
precursors of a dry merger. Furthermore, one could expect that the
effect will not be apparent in the optical spectrum until later on in
the merger.

\begin{figure}
\begin{center}
\includegraphics[width=3.5in]{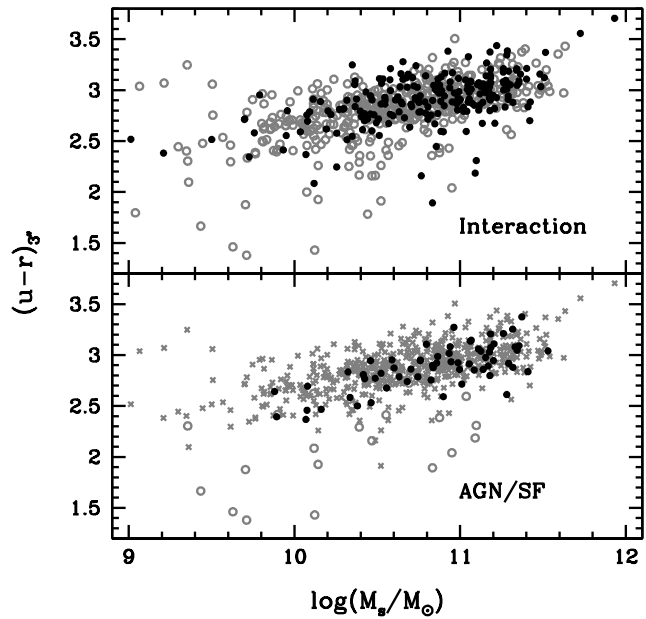}
\end{center}
\caption{Colour-stellar mass relation of our sample. The colours
are extracted from the SDSS DR7 database, dereddened and measured over the
3~arcsec aperture of the fibers used for the spectroscopy.
The sample is divided with respect to visual signs of interaction ({\sl top panel})
or emission line activity ({\sl bottom}). In the top panel, solid black (grey open)
circles correspond to galaxies with (without) visual signs of interaction.
In the bottom panel, solid black (grey open) correspond to galaxies
with AGN (Star-forming) activity. The grey crosses are quiescent galaxies.
}
\label{fig:CMR}
\end{figure}

However, from the histograms on the right-hand side of
figure~\ref{fig:pca1} one can see that the close pairs sample, both
interacting and non-interacting, has a distribution of the $\zeta$
component consistent with the general (i.e. non-close pairs) subsample
of NUV-bright galaxies. This result indicates that elliptical galaxies
in close pairs are more likely to contain small amounts of recent star
formation than a general sample of early-type galaxies. It is
acknowledged that this result could be due to the selection effects
present in \citet{bern06}. However, the dearth of close pair galaxies
at the low end in the distribution of the $\zeta$ component means that
this bias is unlikely. The distribution in the $\eta$ component is not
so useful to discriminate between quiescent spectra and recent star
formation, although the figure shows that the close pair distribution
sits 'roughly' in between the histograms for the NUV bright and faint
galaxies. Given that $\eta$ correlates with colour, one could conclude
that studies purely based on broadband photometry cannot measure these
differences. We note that these distributions are in agreement with 
previous work on early-type galaxies in Hickson Compact Groups, where
a larger scatter towards high values of PC2 (roughly $\zeta$) were
found \citep{pcai}.

In figure~\ref{fig:pca2} we show the PCA components $\eta$ and $\zeta$
as a function of pair separation ({\sl left}); stellar mass ratio ({\sl centre})
and stellar mass ({\sl right}). The black solid (grey open) circles
correspond to the interacting (non-interacting) pairs, and the
vertical lines span the $25^{\rm th}-75^{\rm th}$ percentile of the
distribution in each bin. Out of the three observables, stellar
mass is the only one that correlates significantly with the components, 
appearing redder (i.e. more negative $\eta$) towards higher
masses. The correlation is much weaker with respect to $\zeta$, which
is the proxy for the presence of recent star formation. The mass ratio
does not present any significant trend, and only the non-interacting
pairs give a slight trend towards an increased recent star formation
at small separations, although the scatter of the sample is rather
large.

\subsection{Modelling the Star Formation History}

Since the close pairs sample has a higher mean value of the $\zeta$
component, we expect a large fraction of the sample to have undergone
recent star formation. In order to quantify the effect on the
underlying stellar populations, we explore a two-component star
formation history. The synthetic spectra are generated from the 2007
stellar population models of Charlot \& Bruzual \citep[see
e.g.][]{bc03,bru07}, assuming a \citet{chab} initial mass function.  The
composite model superposes two simple stellar populations: an old
component with age t$_{\rm O}$, allowed to vary between 2 and 14 Gyr,
with the metallicity between $\log(Z/Z_\odot)=-1.5$ and $+0.4$. A
younger component of the same metallicity is added, with age t$_{\rm
Y}$, contributing a mass fraction, f$_{\rm Y}$, which ranges from 0 to
0.5. The age of the young component is taken between 100~Myr and
2~Gyr. Note that the model grid contains a subset of models which are
equivalent to standard simple stellar populations (i.e. f$_{\rm Y}$ =
0), such that a composite model will be chosen only if it improves the
fit.  Also note that the oldest ages considered (14~Gyr) are motivated
by our choice of a standard $\Lambda$CDM cosmology ($\Omega_m=$0.3,
H$_0=70$km/s/Mpc). The final grid of model consists of 65,536 star
formation histories.

Our analysis follows the approach of \citet{bmc}, which involves
multiple age-sensitive spectral features comprising three Balmer lines
(H$\beta$, H$\gamma$ and H$\delta$) and the 4000\AA\ break strength
(D4000), along with a metal-sensitive index, [MgFe] \citep[as defined
in ][]{gonz}. Furthermore, a new definition of equivalent width (EW)
is used, that significantly reduces the age-metallicity degeneracy
over the traditional side-band method \citep[see e.g.][]{sct00}. Our
EWs are based on a new definition of the pseudo-continuum, determined
from a (boosted) median of the surrounding spectra and a 20\AA\
spectral window centered on the line of interest. This definition
has been shown to reduce the contamination of the pseudo-continuum
from neighbouring lines, resulting in a less metal dependent H$\gamma$
and H$\delta$ and a less age dependent [MgFe]. This method also
provides smaller uncertainties in the EW at low S/N \citep{bmc}.

While the BC03 models are not calibrated to accommodate non-solar
abundance ratios, following \citet{yama} we use as a proxy the
comparison between the best fit values of the metallicity when
replacing in the analysis the metal-sensitive index [MgFe] by either
$\langle$Fe$\rangle$ -- defined as 0.5(Fe5270+Fe5335) -- or by the
Mg$b$ index. The difference in the metallicity derived from these two
fits -- labelled as Z$_{\rm Mg}$ and Z$_{\rm Fe}$ -- is given as a
crude estimation of the abundance ratio: $[$Z$_{\rm Mg}/$Z$_{\rm
Fe}]\equiv \log($Z$_{\rm Mg}/$Z$_{\rm Fe})$. The comparison of
observed data and models is done via a standard maximum likelihood
method. The errors for the EWs are estimated from Monte Carlo
realizations of gaussian noise applied to the spectra. These errors
are added in quadrature to the estimated systematic errors for the
models \citep{bc03}.

\subsubsection{Results}

The results of the line strength modelling are shown in figure
\ref{fig:2bst}, as a function of projected separation ({\sl left}); stellar
mass ratio ({\sl centre}) and central velocity dispersion ({\sl right}).  Black
solid (grey open) circles correspond to close pair galaxies classified
as interacting (non-interacting). The error bars give the 
$25^{\rm th}-75^{\rm th}$ percentile range of the distribution
within each bin. From top to bottom, we show the proxy for abundance
ratio ($[Z_{\rm Mg}/Z_{\rm Fe}]$); metallicity; (mass-weighted)
average age, and a quantity that gives the strength of the recent
episode of star formation. The mass and age of the young population is
degenerate, such that a small mass in young stars can be replicated by
a larger mass fraction in older stars. Hence, we parameterise the
effect on the spectra in terms of the ratio between the mass fraction
in the young component to its age (f$_{\rm Y}/$t$_{\rm Y}$).  

Consistently to the PCA studies described above, there is no
significant difference between the stellar populations with respect to
the visual presence of interactions. Furthermore, the populations are
only sensitive to the velocity dispersion, a result already present in
general samples of galaxies \citep[see e.g.][]{bern05}. There is no
significant trend with respect to separation or mass ratio. However,
the modelling of the line strengths do reveal in an independent and
consistent way to PCA that early-type galaxies in close pairs are more
likely to have undergone recent star formation. We find 378 out of 695 galaxies 
(i.e. 54\%) have a significant young population, $\log($f$_{\rm Y}/$t$_{\rm
Y})$ $\geq -2$, which is equivalent to 1\% mass fraction of a 1~Gyr
population.

The correlation between colour and stellar mass can also be used to
understand the connection between pair morphology, emission line
activity and the underlying stellar populations. Figure~\ref{fig:CMR}
shows the ($u-r$) colour-stellar mass diagram, where the colours are
directly obtained from the SDSS DR7 database (de-reddened and measured
within the $3^{\prime\prime}$ aperture of the spectrograph fibers).
In the top panel, solid black (grey open) circles correspond to
galaxies with (without) visual signs of interaction.  In the bottom
panel, solid black (grey open) circles correspond to galaxies with AGN
(star forming) activity. The grey crosses are quiescent galaxies.  As
expected, colour is strongly correlated with mass, which confirms the
trend seen in the PCA components in figure~\ref{fig:pca2}. However,
the departure from the red sequence -- which is an indicator of
younger stellar populations -- does not depend on signs of visual
interaction, agreeing with the PCA result of figure~\ref{fig:pca1}.
However, with respect to emission line activity (bottom panel of
figure~\ref{fig:CMR}), we find a significant trend such that quiescent
galaxies and AGN systems (crosses and solid dots, respectively)
populate the red sequence, whereas galaxies with on-going star
formation are identified as the members of the 'blue cloud'. The
evolution from a weak star forming cloud towards a weak AGN and
subsequently a quiescent galaxy has already been proposed elsewhere
\citep{schw07}. In the conclusion we apply this evolutionary path to
the interpretation of the early-type close pairs.

\begin{figure}
\begin{center}
\includegraphics[width=3.5in]{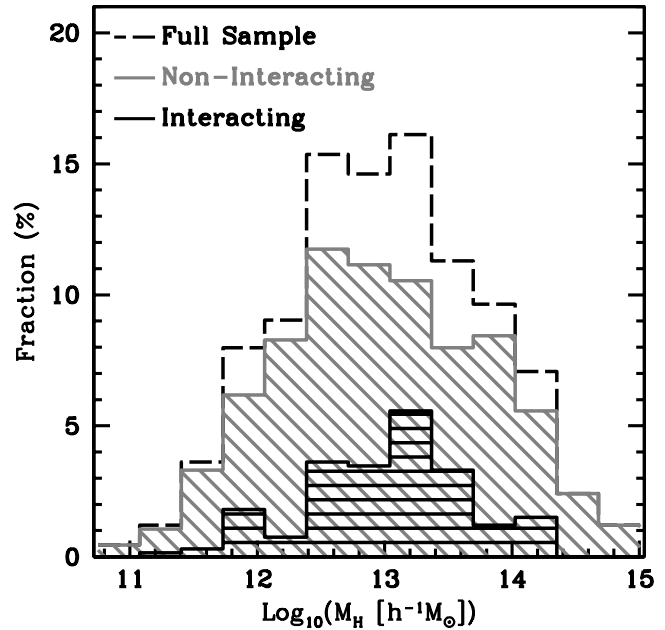}
\end{center}
\caption{The distribution of the sample is shown in terms of 
estimated mass of the dark matter halo of the occupied galaxy group
(M$_{\rm H}$).
The histograms show the distributions of the pairs depending on whether 
the system is visually classified as interacting (black solid) or not 
(grey). The full sample corresponds to the black dashed lines. 
}
\label{fig:ENV1}
\end{figure}

\section{Environment}

We can also explore the properties of the close pairs with
respect to environment. In order to assess the environment of a
galaxy, we use the estimated mass of the host dark matter halo. This
is determined through the group finding algorithm of \citet{yang05},
which identifies galaxy groups starting with a friends-of-friends
algorithm. The membership to these groups follows an iterative process
controlled by the properties of the group and its halo.  The
application of this algorithm to a large sample of SDSS galaxies,
forms the \emph{galaxy groups catalogue} of \citet{yang07}. The
interested reader is directed to those references for an in depth
description of the algorithm and catalogue. A cross correlation of
this catalogue to our close pairs sample reveals a match to 664 of the
695 galaxies. As a measure of environment, we use the halo mass
estimated based on the observed stellar mass of the group.  We note
that a more detailed investigation into the effects of environment as
measured by the mass of the galaxy group halo is also underway on a
much larger sample of ellipticals (Rogers et al. in preparation).

Figure~\ref{fig:ENV1} shows the distribution of our sample in terms of
the host halo mass. The galaxies from our close pairs sample are
mainly located in intermediate mass halos of 
M$_H\sim 10^{13}$M$_{\odot}h^{-1}$. This is in agreement with the
halo mass found by \citet{pasq09} at which galaxies change from being
preferentially star-forming to showing optical-AGN activity.

\begin{figure}
\begin{center}
\includegraphics[width=3.5in]{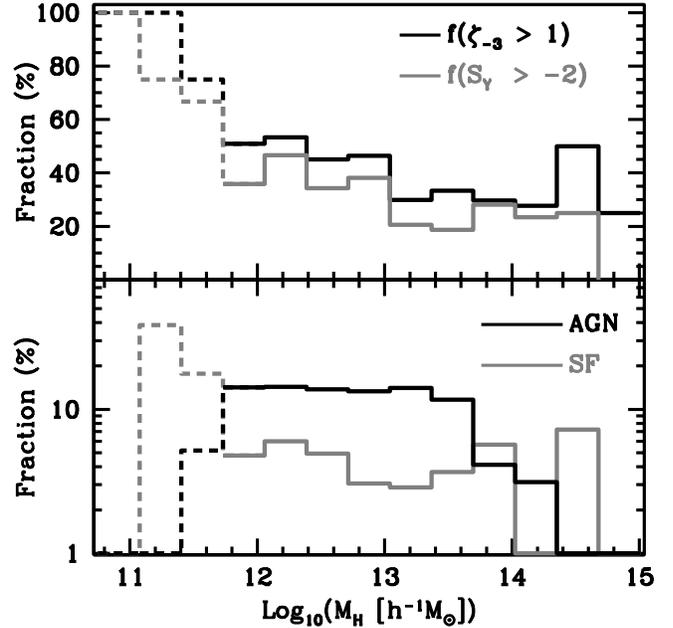}
\end{center}
\caption{
(\emph{Top Panel}) The change in the fraction of galaxies with significant 
recent star formation as a function of the host halo mass 
(M$_{\rm H}$). Shown for both 
PCA and SFH modelling results, where the significant level of RSF is defined 
by $\zeta_{-3}\equiv\zeta/10^{-3}\geq 1$ and ($S_Y$ = ) $\log($f$_{\rm Y}/$t$_{\rm Y})$ $\geq -2$. 
(\emph{Bottom Panel}) The fraction of galaxies containing AGN or star formation 
as a function of environment. The fraction are conditional, $f(AGN, SF | M_H)$, 
refering only to the galaxies contained within the specific halo mass bin.    
The solid grey and black histograms correspond to star-forming 
and AGN galaxies, respectively. The dashed line signify bins 
containing a bias towards low mass galaxies. It is interesting to note the 
drop in AGN fraction after $\log M_H \sim 13.6$  $(M_{\odot} h^{-1})$.
}
\label{fig:ENV2}
\end{figure}

Our sample is clearly too small to map global properties of active
early type galaxies, however it is interesting to see how the activity
of the close pairs are affected. In a manner similar to
\citet{pasq09}, we show in figure~\ref{fig:ENV2} the conditional
fractions of galaxies classified as having undergone significant
recent star formation ({\sl top}) or split with respect to their star
forming / AGN activity ({\sl bottom}).  The conditional fractions are
given by the number of active galaxies within a bin of host halo mass
divided by the total number of galaxies in that bin.  The dashed lines
towards the left of the figure represent the halos for which a
considerable bias is expected towards low mass galaxies as imposed by
the mass of the host halo.  In addition to the limitation of galaxy
mass by the group mass itself, it should be remembered that low mass halos
preferentially host low mass galaxies \citep{yang08}. In a sample of
this size it is difficult to overcome such a bias and this caveat
should be considered alongside the conclusions of this section.

Figure~\ref{fig:ENV2} ({\sl bottom panel}) shows that the fraction of
galaxies classified as AGN is fairly constant\footnote{The small or
zero fraction of AGN at low halo masses is most likely due to the lack
of intermediate and high mass galaxies in these halos. However the
increase in SF in these bins may hint that dust from such activity may
obscure some of the AGN.} up to
M$_H \sim 4\times 10^{13}$M$_{\odot} h^{-1}$,
above which there appears to be a rapid decline. The dearth of AGN
galaxies at high halo masses is consistent with previous
results \citet{kauf04,gil07,pasq09}. The exact reason for the cut off is
not obvious but may be related to the reduction of gas available
due to increased tidal stripping.

The fraction of star forming galaxies is higher in lower mass
halos, although low mass galaxies -- well known to have higher fractions
of star-formation \citep{kauf03} -- dominate within these bins as
mentioned above.  The rest of the sample show a consistent fraction of
star forming galaxies across all environment types. The number of galaxies is
relatively small so a robust conclusion cannot be inferred, but it is
interesting to note that no significant drop in star forming galaxies
is seen with respect to halo mass.

Also shown in figure~\ref{fig:ENV2} ({\sl top panel}), is the effect
on the stellar populations where we consider how the amount of recent
star formation is affected by the mass of the host
halo. We define a ``significant'' amount of recent star formation as 
S$_{\rm Y}\equiv\log($f$_{\rm Y}/$t$_{\rm Y})\geq -2$, or $\zeta$ $\geq 10^{-3}$. 
The fraction of galaxies with RSF drops
by $\sim$10~-~20\% as we move from low/intermediate to high halo masses. We
note that this drop is qualitatively similar, although relatively
higher, than that found in \citet{schw07b}, which may be due to the
high sensitivity of NUV or a function of the enhanced RSF already seen
in the sample.

\section{Conclusions}

We have selected a sample of $\sim$350 close pairs involving {\sl
only} early-type galaxies to measure environment effects. Through the
use of emission line diagnostics, we classify $\sim$ 20\% of the
sample as containing either an AGN or currently undergoing (weak) star
formation. The fraction of these 'active galaxies' is consistent with
an independent sample of early-type systems not in close pairs, taken
from \citet{schw07}.  However, we found an excess of AGN and a lack of
star forming galaxies in close pairs with visible signs of
interaction. This result suggests that during the encounter, galaxies
evolve from a (weak) star forming phase to an AGN phase.  This idea is
supported by the increase both in the AGN fraction and in the luminosity
of the [OIII] (5007\AA ) line towards decreasing pair separation.
Additionally, the specific star formation rate shows a hint of a
decrease with decreasing separation, although the number of star
forming systems is very low (contributing only 4\% to the total sample).

A significant increase is found in the number of early-type galaxies
in close pairs that have undergone a recent star formation episode,
with respect to a control sample.  This is shown both through PCA,
where the close pair sample features high values of the $\zeta$
component (sensitive to young stellar populations), as well as by the
line strength analysis using a grid of 2-component models, where a
large proportion of the sample requires a significant amount of young
stars. These two apparently contradicting scenarios -- namely that the
encounter appears to reduce ongoing star formation and the observation
that the sample has increased levels of recent star formation -- can
be reconciled.

It is speculated here that the observed recent star formation is
triggered during the first phases of the encounter by the interactions
of the outer parts of the galaxies, including the dark matter
halos. Since a large fraction of ellipticals have been found to
contain HI and molecular gas, not only in their interstellar medium,
but also in the form of satellite gas clouds
\citep{knap89,Yng05,mor06,com07,donovan}, the increasing gravitational
perturbations induced by an oncoming neighbour will destabilise these
clouds, driving them towards the galaxy \citep{sof93,dimat}. For instance,
\citet{li} found an enhanced star formation rate up to separations of
100~kpc on a large sample of SDSS galaxy pairs. 
Indeed the accretion of 
gas has been shown to instigate star formation even in 
early-type galaxies \citep[e.g.][]{sof93,pip05,khal08}. The 
simulations of \citet{khal08} suggest that the accretion of gas onto an 
early-type galaxy will be short lived due to the feedback from AGN.

As the pair comes closer -- and within our selection criterion of
R$_{\rm P}<30$~kpc -- the increased gravitational interaction will
enable the removal of angular momentum, driving gas towards the centre
\citep[see e.g.][]{DErc00}. Simulations from \citet{dimat} indicate
that the greatest inflow of material to the centre occurs at
separations $\sim$10~kpc. Hence, at lower separations, most of the
available gas is driven towards the centre, triggering the AGN
activity and possibly quenching star formation.  This is consistent
with the simulations of \citet{joh08}, in which mergers between
elliptical galaxies showed decreasing star formation rates with
progession of the merger and rapid termination at the later stages
coincident with increased black hole accretion. The feedback from the
AGN in such cases should drive out the majority of the gas within the
galaxy into the intergalactic medium. Given the high metallicity of
early type stellar populations we might expect this to contribute to the
abundance of the IGM.

Although speculative, this scenario explains the 
transition seen in the emission and absorption features of our 
sample, and is consistent with the observations of \citet{schw07} 
on a general sample of early-type galaxies. Our results agree 
with the general study of SDSS close pairs of \citet{ell08} who find
that star formation also precedes AGN activity in close pairs of
late-type galaxies.

The environment in which the interaction takes place also seems to affect 
the exact nature of the encounter. In low mass halos the initiated 
star formation appear to continue for longer into the interaction, i.e. 
to appear within our selection window of 30 kpc separation. The amount of recent 
star formation also appears to be higher in these low mass halos (possibly as a 
consequence) as well as in intermediate mass halos, which harbour most of the 
AGN activity. In contrast, halo masses greater than M$_H\simgt 4\times 10^{13}$
M$_\odot h^{-1}$ appear 
to have less AGN activity (almost none) and reduced amounts of RSF.

\section*{Acknowledgements}
SK gratefully acknowledges a Research Fellowship from the Royal
Commission for the Exhibition of 1851, a Senior Research Fellowship
from Worcester College, Oxford and support from the BIPAC Institute at
Oxford. This work has made use of the delos computer cluster in the
physics department at King's College London. Funding for the SDSS and
SDSS-II has been provided by the Alfred P. Sloan Foundation, the
Participat- ing Institutions, the National Science Foundation, the US
Department of Energy, the National Aeronautics and Space
Administration, the Japanese Monbukagakusho, the Max Planck Society
and the Higher Education Funding Council for England. The SDSS website
is http://www.sdss.org/. The SDSS is managed by the Astrophysical
Research Consortium for the Participating Institutions.


\end{document}